\documentclass{PoS}
\usepackage{epsfig}

\newcommand{\ave}[1]{\big\langle #1 \big\rangle}

\PoS{PoS(BDMH2004)097}

\title{The galaxy-dark matter bias in the Garching-Bonn Deep Survey}
\ShortTitle{Galaxy bias in GaBoDS}

\author{\speaker{Patrick Simon}$^1$, Peter Schneider$^1$, Thomas Erben$^1$,
  Mischa Schirmer$^2$, \newline Christian Wolf$^3$ and Klaus Meisenheimer$^4$\\
  $^1$ Institut f\"ur Astrophysik und Extraterrestrische Forschung,
  Auf dem H\"ugel 71, 53121 Bonn, Germany\\
  e-mail: \email{psimon@astro.uni-bonn.de}\\
  $^2$ Isaac Newton Group of Telescopes, Santa Cruz de La Palma, Espa\~{n}a\\
  $^3$ Department of Physics, Denys Wilkinson Bldg., University of
  Oxford, Keble Road, Oxford, UK\\
  $^4$ Max-Planck-Institut f\"ur Astronomie, K\"onigsstuhl 17, 69117
  Heidelberg, Germany}

\abstract{We measured the bias and correlation factor of galaxies with
  respect to the dark matter using the aperture statistics including
  the aperture mass from weak gravitational lensing. The analysis was
  performed for three galaxy samples selected by R-band magnitudes;
  the median redshifts of the samples were $\ave{z}=0.34,0.49$ and
  $0.65$, respectively. The brightest sample has the strongest peak in
  redshift and can therefore be most accurately represented by a
  single redshift. The data used is the GaBoDS, and the COMBO-17
  survey for an accurate estimate of the redshift distribution of the
  galaxies.  Assuming the currently favoured $\Lambda\rm CDM$ model as
  cosmology, we obtained values for the linear stochastic galaxy-dark
  matter bias on angular scales $1^\prime\le\theta\le20^\prime$.  At
  $10^\prime$, the median redshifts of the samples correspond to a
  typical physical scale of $0.90,1.25,1.56~\rm Mpc/h$ with $h=0.7$,
  respectively.

  Over the investigated range of physical scales the bias factor $b$
  stays almost constant, possibly rising on the smallest scales.  Here
  the errors are largest, however. Averaging the measurements for the
  bias over the range $4^\prime\le \theta_{\rm ap}\le 18^\prime$,
  weighting with the cosmic variance error, yields $b=0.89(5),
  0.79(5), 0.89(5)$, respectively ($1\sigma$).  Galaxies are thus less
  clustered than the total matter on that particular range of scales
  (anti-biased).  This is what also has been observed by Hoekstra et
  al. (2002), albeit that their increasing trend towards a larger bias
  factor on larger scales is not visible in our analysis. As for the
  correlation factor $r$ we see, as Hoekstra et al., a slight increase
  to $r=1$ in the last angular bin from an almost constant value on
  smaller scales; the weighted average here over the same range as
  before is $r=0.8(1), 0.8(1), 0.5(1)$, respectively.  Therefore, on
  these scales we find a degree of stochasticity or/and nonlinearity
  in the relation between dark matter and galaxy density.  Within the
  measurement uncertainties and over the redshift range represented by
  our galaxy samples we do not see an evolution with redshift of the
  bias. }

\FullConference{Baryons in Dark Matter Halos\\
                 5-9 October 2004\\
                 Novigrad, Croatia}

\begin{document}

\section{Introduction}

The current paradigm of the $\Lambda\rm CDM$ model is mainly supported
by observations of the cosmic microwave background, the apparent
luminosity-distance relation, the element ratios in the primordial gas
and the ages of the oldest objects. The most recent surprising
discovery is the dark energy ($\Omega_\Lambda\simeq0.7$). According to the
model the structure in the matter distribution formed from an almost
homogeneous state with primordial adiabatic fluctuations by
gravitational collapse in a globally expanding Universe.  Mainly
responsible for the collapse is a collisionless, pressureless kind of
matter with a small velocity dispersion, the cold dark matter
($\Omega_{\rm m}\simeq 0.3$); it reveals itself only indirectly
through its effect on the baryonic matter ($\Omega_{\rm b}\simeq
0.05$) by its gravitational field: for instance in the dynamics of
galaxies, galaxy rotation curves, the distribution of x-ray emitting
gas in galaxy clusters or gravitational lensing.

In this picture, galaxies take only a minor part, in the sense that
their mass is completely negligible compared to the total mass
content. The laws determining the variety of galaxy types and their
distribution are very complex.  It would be very surprising if this
complexity would eventually have resulted in a simple linear
one-to-one relationship between the galaxy density and total matter
density making the galaxies so-called unbiased tracers of the total
matter distribution.  On the other hand, if we knew the differences in
the distribution of galaxies with respect to the dark matter as a
function of galaxy type and redshift we could learn more about the
galaxy formation and evolution process.

Biasing between two density fields, say $\rho_i$ for the dark matter
density and $\rho_j$ for the galaxy number density, can in general be
quantified invoking the joint probability distribution function (pdf),
$P\left(\delta_i,\delta_j\right)$, of the density contrasts,
$\delta=\rho/\ave{\rho}-1$, at the same point in the density fields at
some redshift (local Eulerian bias). Using second order statistics
only, we can distinguish between two possibly scale-dependent bias
parameters $b_{ij}=\sqrt{\ave{\delta_i^2}/{\ave{\delta_j^2}}}$ and
$r_{ij}=\ave{\delta_i\delta_j}/{\sqrt{\ave{\delta_i^2}\ave{\delta_j^2}}}$,
which we call the \emph{linear stochastic bias}. The quantity $b_{ij}$
is a measure of the difference in clustering while $r_{ij}$ measures
the stochasticity and non-linearity in the relation between $\delta_i$
and $\delta_j$. A more advanced scheme that also separates
non-linearity and stochasticity has been proposed by Dekel~\&~Lavav
(1999). This scheme is, however, not applicable if only second order
statistics are used as in the technique applied in this paper.

As we know from observations (cf. Simon (2004), introduction therein)
galaxy biasing is a function of smoothing scale, redshift, galaxy type
and luminosity. As some examples, on scales larger than about $10~\rm
Mpc$ and at low redshifts, galaxies as a whole are not biased with
respect to the dark matter but slightly anti-biased, $b<1$, on
intermediate scales. The large-scale bias increases towards larger
redshifts, and red and blue galaxies are differently biased.

In this paper, we apply the method of Hoekstra et al. (2002) (Sect.
\ref{methodsection}) to the Garching-Bonn Deep Survey (Sect.
\ref{gabodssection}) to obtain the linear stochastic biasing
coefficients, $r$ and $b$, of three galaxy samples, binned by their
apparent R-band magnitude, with respect to the total matter
distribution. The final result is presented in Sect.
\ref{resultssection}.

\section{Method}
\label{methodsection}

The method we are applying here, based on the work of Schneider (1998)
and van Waerbeke (1998), is explained in more detail Hoekstra et al.
(2002). This technique uses the weak gravitational lensing effect to
map the total matter distribution along the line-of-sight; the tidal
gravitational field of the matter inhomogeneities is imprinted in the
coherent shape distortions of distant \emph{background} galaxies.
Considering the tangential alignments of these galaxies about some
aperture centre, the aperture mass $M_{\rm ap}\left(\theta_{\rm
    ap}\right)$ gives a noisy measure for the projected line-of-sight
dark matter density contrast smoothed to some typical scale. The scale
depends on the aperture filter and aperture size $\theta_{\rm ap}$.
Employing the same kind of aperture statistics also provides the
aperture number count $N\left(\theta_{\rm ap}\right)$ which is the
projected number density contrast of \emph{foreground} galaxies.
Averaging over many apertures results in an estimate for the linear
stochastic bias between galaxies and dark matter
\begin{equation}
  b\left(\theta_{\rm ap}\right)=f_1\,\,
  \sqrt{\frac{\ave{N^2\left(\theta_{\rm ap}\right)}}{\ave{M_{\rm
          ap}^2\left(\theta_{\rm ap}\right)}}} ~~;~~ 
  r\left(\theta_{\rm ap}\right)=f_2\,\,
  \frac{\ave{N\left(\theta_{\rm ap}\right)M_{\rm ap}\left(\theta_{\rm
          ap}\right)}}{\sqrt{\ave{N^2\left(\theta_{\rm
            ap}\right)}\ave{M^2_{\rm ap}\left(\theta_{\rm ap}\right)}}}
  \; .
\end{equation}
$f_1$ and $f_2$ are calibration factors computed based on the
cosmological model and the redshift distribution of the background and
foreground galaxies.
In practise, the second order moments $\ave{N^n\left(\theta_{\rm
      ap}\right)M^m_{\rm ap}\left(\theta_{\rm ap}\right)}$ with
$n+m\le2$ are obtained as integral transforms of observed two-point
correlation functions: the angular correlation of the foreground
galaxies ($n=2$), galaxy-galaxy lensing ($n=m=1$) and cosmic shear
correlations ($m=2$).

\section{GaBoDS: The Garching-Bonn Deep Survey}
\label{gabodssection}

The GaBoDS (Schirmer et al. 2003) comprises roughly $20~\rm deg^2$ of
high-quality data (seeing better than one arcsec) in R-band taken with
the Wide Field Imager (WFI) at the 2.2m telescope of MPG/ESO at La
Silla, Chile; the $34^\prime\times 33^\prime$ field of view is covered
with 8 CCD chips.  The data set was compiled by Mischa Schirmer and
Thomas Erben partly from archived ESO data, partly from new
observations ($3.7~\rm deg^2$). They selected and reduced the data
especially for applications with focus on weak gravitational lensing.
The data set can roughly be categorised into a shallow ($t\le
7\rm~ksec$, $9.6\rm~deg^2$), medium ($7~{\rm ksec}<t\le10~{\rm ksec}$,
$7.4~\rm deg^2$) and deep ($10~{\rm ksec}<t\le 56\rm~ksec$, $2.6
\rm~deg^2$) set depending on the total number of frames usable for the
co-addition of each field.  This work uses only the deep and medium
deep category, in total consisting of $28$ WFI fields.

We subdivided the galaxy catalogues into three foreground (FORE-I,
FORE-II, FORE-III) and one background bin (BACK, for lensing)
representing different median redshifts:
\begin{center}
\begin{tabular}{llcc}  
  sample&bin limits [mag]&\#objects&$\ave{z}\pm(\Delta z)^2$\\ \hline\\
  FORE-I & $19.5\le{\rm  R}<21.0$ & $2.5\times 10^4$ & $0.34\pm0.09$  \\ 
  FORE-II& $21.5\le{\rm  R}<22.0$  &  $4.1\times 10^4$  &  $0.49\pm0.12$\\ 
  FORE-III & $22.5\le{\rm  R}<23.0$   &  $9.5\times 10^4$   &  $0.65\pm0.14$\\ 
  BACK & $21.5\le{\rm  R}<24.0$   &  $3.2\times 10^5$  &  $0.67\pm0.15$
\end{tabular}
\end{center}
Three of the $28$ patches were part of the COMBO-17 survey (Wolf et
al. 2004). They provide, among other things, highly accurate
photometric redshifts of galaxies brighter than $24~\rm mag$ with an
uncertainty of roughly $\delta z\le 0.01(1+z)$; see Fig.
\ref{biasfig}.  The average redshift distribution in these fields was
used as an estimate for the distribution of all galaxies in the GaBoDS
sample.

\section{Results}
\label{resultssection}

\begin{figure}
  \begin{center}
    \epsfig{file=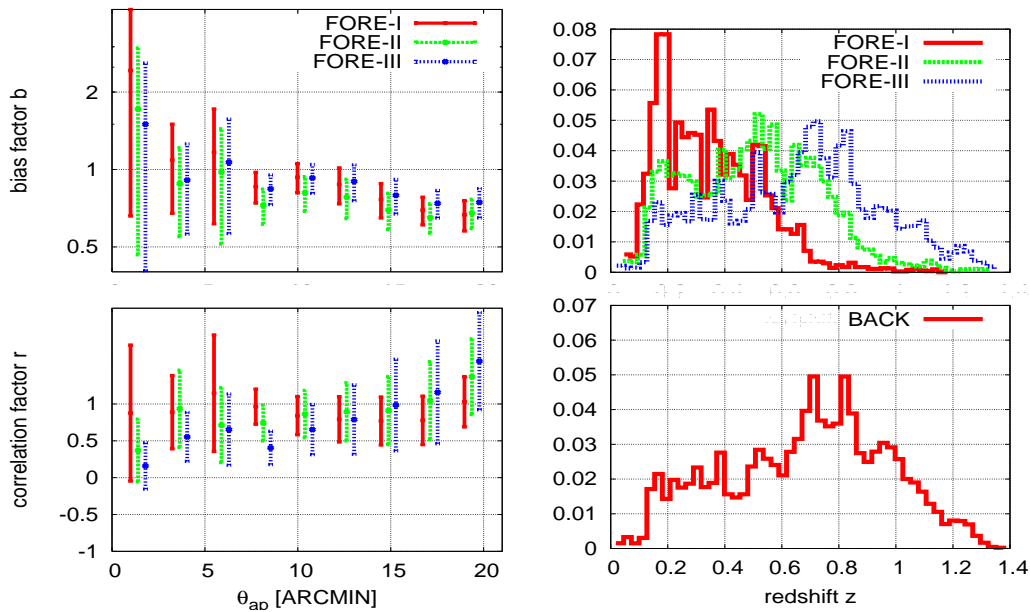,width=14cm,angle=0}
  \end{center}
  \caption{Linear stochastic bias as observed for three foreground
    galaxy magnitude bins (upper left: bias factor, lower left:
    correlation factor) as a function of aperture radius; $\theta_{\rm
      ap}=10^\prime$ corresponds to $0.90,1.25,1.56~\rm Mpc/h$ for
    FORE-I, FORE-II and FORE-III, respectively, taking into account
    the aperture filter properties. The distribution in redshift of
    the galaxy samples (from COMBO-17) is plotted in the upper right
    panel (foreground) and the lower right panel (background, used for
    lensing). The bias parameters have been calibrated assuming a
    fiducial cosmology: $\Omega_{\rm m}=0.3, \Omega_\Lambda=0.7,
    \sigma_8=0.9, \Gamma=0.21, h=0.7$ and a transfer function for
    adiabatic CDM.\label{biasfig}}
\end{figure}

The final result of our effort is comprised in Fig. \ref{biasfig}.
Within the measurement uncertainties ($1\sigma$ cosmic variance) the
bias parameters stay approximately constant over the probed physical
scales, maybe rising towards larger scales, for $r\left(\theta_{\rm
    ap}\right)$, or smaller scales, for $b\left(\theta_{\rm
    ap}\right)$.  The results for FORE-I are comparable to Hoekstra et
al.  (2002).  Considering the uncertainties and the mean redshifts of
the three galaxy foreground samples the bias evolution on these scales
has to be smaller than $\Delta b\le0.1$ and $\Delta r\le 0.2$
($2\sigma$ confidence) for $0.3\lesssim z\lesssim 0.65$.


\end{document}